\documentclass[%
 reprint,
 amsmath,amssymb,
 aps,
 pra,
]{revtex4-2}

\usepackage{graphicx}
\usepackage{dcolumn}
\usepackage{bm}
\usepackage{hyperref}
\usepackage{xcolor}

\begin{document}
\preprint{APS/123-QED}

\title{$\beta$-Variational Autoencoder as an Entanglement Classifier}

\author{Nahum S\'a}
 \email{nahumsa@cbpf.br}
\author{Itzhak Roditi}%
 \email{roditi@cbpf.br}
 \altaffiliation[Also at ]{Institute for Theoretical Physics, ETH Zurich 8093, Switzerland.}
\affiliation{
 Centro Brasileiro de Pesquisas F\'isicas, \\ 
 Rua Dr. Xavier Sigaud 150, 22290-180 Rio de Janeiro, Brazil
}

\noaffiliation

\date{\today}

\begin{abstract}
We focus on using an architecture similar to the $\beta$-Variational Autoencoder ($\beta$-VAE) to discriminate if a quantum state is entangled or separable based on measurements. We split the data into two sets, the set of local and correlated measurements. Using the latent space, which is a low dimensional representation of the data, we show that restricting ourselves to the set of local data it is not possible to distinguish between entangled and separable states. Meanwhile, when considering both correlated and local measurements, an accuracy of over 80\% is attained in the structure of the latent space.
\end{abstract}

\maketitle


\section{Introduction}
\label{Sec: Introduction}

Entanglement is one of the most outstanding properties in quantum systems. It introduces correlations that are non-classical and may occur between otherwise non-interacting systems. Its usefulness emerges in such areas as quantum information, quantum computation, quantum cryptography and quantum metrology. It is also crucial to the phenomena of quantum teleportation~\cite{BB_+_quantumtelep}.
Due to the importance of entanglement, which appears in so many instances, it is no wonder that there is a huge interest in finding methods that can detect, classify and quantify it \cite{amico2008entanglement,ma2018ent, horodecki2009quantum, plenio2014introent}. 

On the other hand, deep learning (DL) techniques are becoming one of the most important assets in the physicists' toolbox, for instance helping understanding patterns that have little or no bias from a previously established theoretical framework. In general, DL techniques have been used in relation to computer science, and some successful examples are the technologies associated with pattern recognition, especially computer vision \cite{krizhevsky2012imagenet,sermanet2013vision}, as well as some applications that received attention from the media, like Alpha GO~\cite{silver2017GO}. Lately this kind of approach found its way in physics, going beyond computer science. Some recent applications emerge in condensed matter physics \cite{carrasquilla2017machine}, quantum many-body physics  \cite{carleo2017solving, torlai2018neural} and molecular modeling \cite{rupp2012molecular}. DL techniques are even being applied as a way to unveil how physical concepts emerge \cite{ETH1, ETH2}.

In this paper, we are concerned to the problem on how to distinguish between entangled and separable states, this is known to be an NP-hard problem \cite{gurvits2003classical}, therefore there is no known classical algorithm that could solve this problem efficiently. 

Specifically, we analyze a method to encode the high dimensional labeled data coming from measurements of 2 qubit states, that corresponds to density matrices with 15 parameters, into a lower-dimensional representation which we call latent space. We deal with this problem through a Neural Network architecture that is similar to the so called a $\beta$-Variational Autoencoder~\cite{higgins2017BVAE} which is used as a tool to distinguish between entangled and separable states.

In section~\ref{Sec: Data} we explain how we simulated and labelled the data. In section~\ref{Sec: Model} we detail how to use the $\beta$-Variational Autoencoder architecture for our problem and specify the loss function used. We discuss and show the results of the model in section~\ref{Sec:Results + Discussion}, and we finish with our conclusions and future works on section~\ref{Sec: Conclusion}.

\section{Data}
\label{Sec: Data}

There are several methods to distinguish between entangled and separable states, as summarized in Horodecki et al.~\cite{horodecki2009quantum}. Here we chose to study the case of bipartite entanglement of 2 qubit states, applying the Positive Partial Trace (PPT) criterion (sometimes called Perez-Horodecki criterion) that was proposed first by Perez \cite{Perez_PPT} and extended by Horodecki et al. \cite{HORODECKI_PPT} where it was shown that it provides a sufficient and necessary condition for a two-qubit system to be entangled.

In Quantum Theory the most general way to describe a quantum state is using a density matrix which is represented by a linear operator having the following properties: (1) Unity trace ; (2) Positivity. 

The PPT Criterion consists of using the partial transpose transformation on a density matrix. If after the transformation, the state of two qubits is completely positive, then it is separable (SEP). Otherwise, it is entangled (ENT). Considering $\rho \in C^2 \otimes C^2$ as a density matrix of two qubits and T as the transpose map, the PPT can be synthesized by the following expression:

\begin{equation}
\begin{split}
    (I\otimes T)\rho \geq 0 \Rightarrow \rho \in \mathrm{SEP} \\
    (I\otimes T)\rho < 0 \Rightarrow \rho \in \mathrm{ENT}
\end{split}
\end{equation}

Since we are going to use supervised learning we need labeled data. To generate the data we used Qutip \cite{qutip2012,qutip2013} to simulate random density matrices and use the PPT criterion to label the data, we chose the label "1" for entangled states and the label "0" for separable states. These labels are one-hot encoded for the input of the Neural Network. We then measure on the Pauli matrices basis $\{\sigma_i \otimes \sigma_j \}$, where $i,j \in {0,x,y,z}$ and $\sigma_0 = I$. All measurements are labeled using the following convention:

\begin{equation}
    M_{ij} = \mathrm{Tr}\big[\rho(\sigma_i \otimes \sigma_j)\big]
\end{equation}

For the two-qubit case we need 15 measurements, excluding $M_{00}$ that always equals to 1 because of the density matrix definition, in order to have a complete tomography of the state. One can split the tomographic-complete measurements into two disjoint sets: correlated measurements, $M_{ij}$ such that $i,j \neq 0$, having 9 measurements and local measurements, $M_{ij}$ such that $i=0$ or $j=0$, having 6 measurements.

Using this convention, we have three types of training and validation data, each with 5000 and 3000 samples, respectively. For convenience, we call these three types: the tomographic-complete dataset, correlated measurements dataset and local measurements dataset.  The data used is slightly unbalanced, because we have approximately $65\%$ of the states belonging to the entangled class. If we choose a classifier that takes into account only the most frequent class we get an accuracy of $65\%$, this "dummy" classifier will be our baseline model.

\section{Model}
\label{Sec: Model}

\begin{figure}
    \centering
    \includegraphics[width=25em]{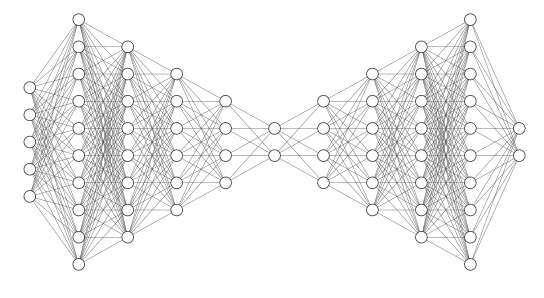}
    \caption{The architecture used on this paper which resembles a VAE (or $\beta$-VAE) architecture, which consists of an encoder and a decoder with a probabilistic latent space. The nodes represent Neurons of the neural Network and the edges represent connectivity between nodes, which is assumed to be completely connected. In this paper, only the input and output depends on the dataset, all others are independent. The encoding structure is of size 128, 64, and 32. The latent space is of size 2, and the decoder structure is of size 32, 64, 128 and the output of size 2.}
    \label{fig: VAE}
\end{figure}

The Variational Autoencoder (VAE) was proposed by Diederik and Welling~\cite{VAE_Paper} and is most commonly used for generative modeling. It has been extended by Higgins et al.~\cite{higgins2017BVAE} for an architecture which is called $\beta$-Variational Autoencoder ($\beta$-VAE) which creates unraveled (disentangled) representations on the latent space (usually a lower-dimensional representation of the data).  Both models can be represented by the same graph shown in figure~\ref{fig: VAE}.

The main principle beneath the VAE, or $\beta$-VAE, is the use of a probabilistic latent space, which is a lower-dimensional representation of the data, that, as assumed, follows some prior distribution. The most common choice is the Gaussian distribution $\mathcal{N}(0,1)$ which will be used in this work. 

Our main result is that we can encode the high dimensional input of measurements into a two-dimensional latent space that acts as a classifier of entanglement using an architecture that resembles a $\beta$-VAE. This representation cannot be obtained on a Feed-Forward Neural Network because it would create a discontinuous latent space. Besides using the latent space as a classifier, we use it to discriminate between the kinds of measurements being made, thus showing that local measurements are not able to characterize if a state is entangled or separable.

We trained from end-to-end using back propagation with a two-component loss function that consists of a categorization loss and a latent loss. Our choice of categorization loss $\mathcal{L}_\mathrm{cat}(\mathbf{y},\mathbf{\hat{y}})$ is the categorical cross-entropy, where $y_i$ is the true label and $\hat{y}_i$ is the predicted label. For the latent loss $\mathcal{L}_\mathrm{KL}(\mu,\sigma)$ we chose the Kullback-Leibler (KL) Divergence. The total loss is given by the sum of these two losses:

\begin{equation}
    \label{Eq: Loss}
    \begin{split}
    \mathcal{L}_{\mathrm{total}}  = r_{cat} & \overbrace{\bigg[\sum_i y_{i} \mathrm{log}(\hat{y}_{i}) \bigg]}^{\mathcal{L}_\mathrm{cat}(\mathbf{y},\mathbf{\hat{y}})}  \\ & + \beta \overbrace{\bigg[ \frac{1}{2}\sum_i (\sigma_i + \mu_i^2 -1 + \mathrm{log}(\sigma_i) \bigg]}^{\mathcal{L}_\mathrm{KL}(\mathbf{\mu},\mathbf{\sigma})}
    \end{split}
\end{equation}

Where $r_{cat}$ and $\beta$ are weighting coefficients which are hyperparameters of our model and should be optimized for our task.

For the training, we adopt the Adam algorithm~\cite{kingma_Adam} as the optimizer with the `reduce learning rate on plateau' method callback on Keras framework \cite{chollet2015keras}. In all layers, except the last layer, we used LeakyReLU Activation in conjunction with a Dropout layer \cite{srivastava2014dropout} to avoid overfitting. For the last layer, our choice was the softmax activation in order to capture the probability of the state being separable or entangled.

We trained the model for 100 epochs, starting with learning rate 0.005, batch size of 256, and using $r_{cat} = 500$ and $\beta = 1$ for each data set.  In order to find those hyperparameters we discuss the  methods for hyperparameter tuning on sec~\ref{Sec:Results + Discussion}.

\section{Results and Discussion}
\label{Sec:Results + Discussion}

\begin{figure}
    \centering
    \includegraphics[width=27em]{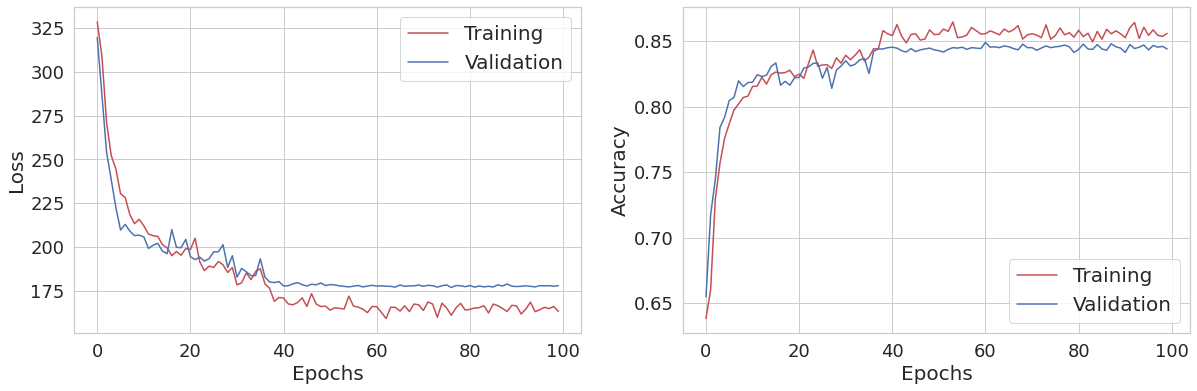}
    \caption{The plot of the Loss function (left) and accuracy of the model (right) during training and evaluation on unseen data. The behavior of the loss function is the same for all datasets but with different accuracy, as specified in the text.}
    \label{fig: Loss}
\end{figure}

\begin{figure}
    \centering
    \includegraphics[width=27em]{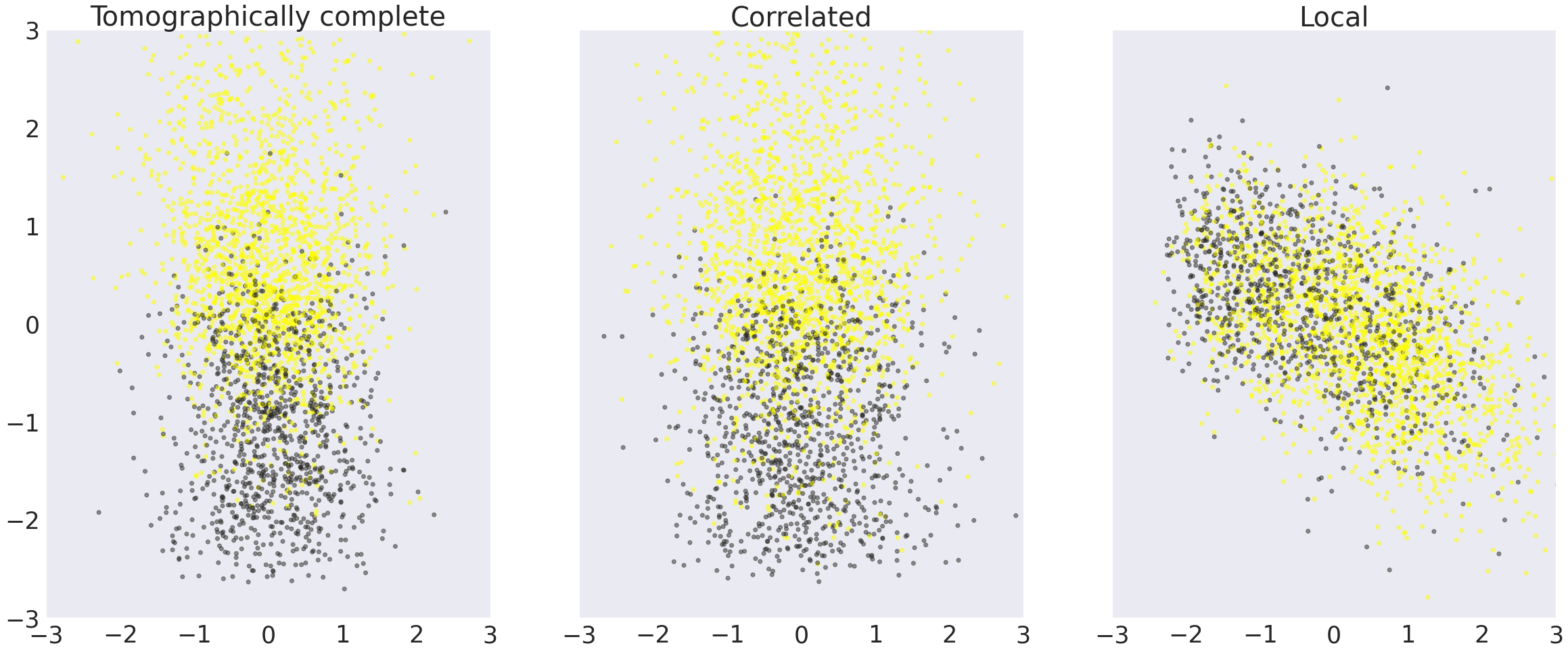}
    \caption{The plot of the latent space for each validation dataset (examples that weren't previously seen by the model). Yellow dots represent entangled states and black dots represents separable states. From left to right tomographic-complete measurements, correlated measurements and local measurements.
    We see that there is clustering on both tomographic-complete measurements and correlated measurements, but we find no clustering on local measurements, showing that the importance of each measurement for entanglement detection is different, therefore we could use only the correlated measurements for detecting entanglement. }
    \label{fig: Latent Space all}
\end{figure}

We trained and evaluated our model for the three datasets, as specified in sections~\ref{Sec: Data} and \ref{Sec: Model}. In our model, we will encode the information of the 15-dimensional input into a 2 dimensional latent space as represented in figure~\ref{fig: Latent Space all}, in which is possible to see that there are correlations between the dataset used and the latent space. The loss, for the tomographically-complete set, regarding training is showed in figure~\ref{fig: Loss}. For the other datasets, the loss behaves similarly when varying the accuracy.

The latent space is divided into two types of distributions, one regarding the separable states and the other regarding entangled states. These distributions can be distinguished by a line that depends on the initialization of the weights, due to the random initialization the line can be on the X or Y axis. Thus, in order to find which one suits better, we can plot the latent space for the known training data and choose the suited line that cuts the plane to distinguish between separable and entangled states.

For tomographic-complete measurements, we see a clear distinction between entangled and separable states, therefore we can use the latent space as an entanglement classifier. For instance, if we chose all points with $y>0$ to be separable we find an accuracy of 80\%, comparing to the accuracy of the whole model, which is 84\%. Therefore, to use only the latent space is effective as a discriminator of entanglement. 

The same can be done for correlated measurements. Indeed, choosing $y>0$ to be separable states we find an accuracy of 80\% and for the full model, we find an accuracy of 83\%. On the other hand, for local measurements choosing $y>0$ on the latent space gives approximately the same accuracy as the whole model, namely, 63\%. 

It is interesting to note, as stated before, that the latent space representation depends on the type of measurements being made. For the correlated measurements ($M_{ij}$ such that $i,j \neq 0$) we see that the latent space still clusters into two different classes just as the case where we use tomographically-complete measurements. On the other hand, for local measurements($M_{ij}$ such that $i=0$ or $j=0$) it is not possible to distinguish between separable and entangled states using the latent space. This is a feature that becomes evident when using the architecture that resembles a $\beta$-VAE.

\begin{figure}
    \centering
    \includegraphics[width=25em]{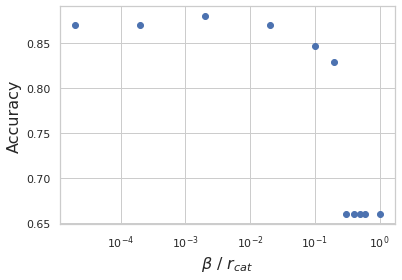}
    \caption{Plot showing the $\beta / r_{cat}$-dependency of the accuracy for tomographic-complete measurements. There are two regimes that are defined by the ratio between $\beta$ and $r_{cat}$, this happens because the $\beta$ factor enforces the latent space distribution for being equal to a Gaussian distribution $\mathcal{N}(0,1)$ if the $\beta$ is considerably smaller than $r_{cat}$ the latent space distribution doesn't need to be Gaussian. }
    \label{fig: Accuracy/Beta}
\end{figure}

To evaluate the hyperparameters of the model we varied the $\beta$ factor of the loss equation (Eq.~\ref{Eq: Loss}) as shown in figure~\ref{fig: Accuracy/Beta}. As can be seen, when $\beta / r_{cat} > 0.3$ the accuracy of the model goes down considerably. 

As expected, the $\beta$ factor multiplying the KL Divergence forces the latent space distribution to the same  prior Gaussian $\mathcal{N}(0,1)$ distribution. Analyzing the shape of our result distribution, mainly the entangled states, we see that it is not Gaussian at all, therefore enforcing the KL Divergence condition will diminish the accuracy of the model.

\section{Conclusion}
\label{Sec: Conclusion}

In this paper, we propose a novel way to use the latent space of a $\beta$-Variational Autoencoder to encode the information concerning the entanglement of the quantum state using a set of tomographically-complete measurements.

We divide this set into two disjoint sets, one for correlated measurements and the other for local measurements in order to analyze if there is any difference between these two types of measurements. 

Applying our method on a tomographically-complete set of measurements of two-qubit system, we can distinguish between entangled and separable states with high precision both in the prediction of the model (84\%) and using the latent space as an entanglement classifier (83\%). In addition, for correlated measurements, of type $\sigma_{x,y,z} \otimes \sigma_{x,y,z}$, we also can distinguish between entangled and separable states, but with less precision for the whole model (82\%) and using only the latent space (80\%) compared to the set of tomographically-complete measurements. 

On the other hand, applying for local measurements the model is not able to learn any representation of entanglement in the latent space, showing that local measurements of type $\sigma_{x,y,z} \otimes I$ or $I \otimes \sigma_{x,y,z}$ are not able to characterize if the state is entangled or separable.

This result is supported by quantum theory because entangled states are expected to show non-locality given by correlated measurements, on the other hand, separable states are expected to be characterized by local measurements. Our model provides a novel way to identify local and correlated measurements, by using the latent space.

In the future, we intend to analyze if this type of model can find accurate description of bipartite or multipartite entanglement for more than two qubits.

\begin{acknowledgments}
We would like to thank Ivan Oliveira, João Ribeiro, Luiz Bispo, and Leonardo Cirto for useful comments on the manuscript. I. R. thanks the Quantum Information Group of ETH-Zurich for their warm hospitality. This study was financed in part by the Coordena\c c\~ao de Aperfei\c coamento de Pessoal de N\'ivel Superior – Brasil (CAPES) – Finance Code 001.

The source code is available on GitHub: \href{https://github.com/nahumsa/Entanglement-VAE}{https://github.com/nahumsa/Entanglement-VAE}.
\end{acknowledgments}

\bibliographystyle{plain}
\bibliography{Bibliography.bib}

\begin{thebibliography}{10}

\bibitem{BB_+_quantumtelep}
Charles~H. Bennett, Gilles Brassard, Claude Cr\'epeau, Richard Jozsa, Asher
  Peres, and William~K. Wootters.
\newblock Teleporting an unknown quantum state via dual classical and
  einstein-podolsky-rosen channels.
\newblock {\em Phys. Rev. Lett.}, 70:1895--1899, Mar 1993.

\bibitem{amico2008entanglement}
Luigi Amico, Rosario Fazio, Andreas Osterloh, and Vlatko Vedral.
\newblock Entanglement in many-body systems.
\newblock {\em Reviews of modern physics}, 80(2):517, 2008.

\bibitem{ma2018ent}
Yue-Chi Ma and Man-Hong Yung.
\newblock Transforming bell’s inequalities into state classifiers with
  machine learning.
\newblock {\em npj Quantum Information}, 4(1):1--10, 2018.

\bibitem{horodecki2009quantum}
Ryszard Horodecki, Pawe{\l} Horodecki, Micha{\l} Horodecki, and Karol
  Horodecki.
\newblock Quantum entanglement.
\newblock {\em Reviews of modern physics}, 81(2):865, 2009.

\bibitem{plenio2014introent}
Martin~B Plenio and Shashank~S Virmani.
\newblock An introduction to entanglement theory.
\newblock In {\em Quantum Information and Coherence}, pages 173--209. Springer,
  2014.

\bibitem{krizhevsky2012imagenet}
Alex Krizhevsky, Ilya Sutskever, and Geoffrey~E Hinton.
\newblock Imagenet classification with deep convolutional neural networks.
\newblock In {\em Advances in neural information processing systems}, pages
  1097--1105, 2012.

\bibitem{sermanet2013vision}
Pierre Sermanet, Koray Kavukcuoglu, Soumith Chintala, and Yann LeCun.
\newblock Pedestrian detection with unsupervised multi-stage feature learning.
\newblock In {\em Proceedings of the IEEE conference on computer vision and
  pattern recognition}, pages 3626--3633, 2013.

\bibitem{silver2017GO}
David Silver, Julian Schrittwieser, Karen Simonyan, Ioannis Antonoglou, Aja
  Huang, Arthur Guez, Thomas Hubert, Lucas Baker, Matthew Lai, Adrian Bolton,
  et~al.
\newblock Mastering the game of go without human knowledge.
\newblock {\em Nature}, 550(7676):354--359, 2017.

\bibitem{carrasquilla2017machine}
Juan Carrasquilla and Roger~G Melko.
\newblock Machine learning phases of matter.
\newblock {\em Nature Physics}, 13(5):431--434, 2017.

\bibitem{carleo2017solving}
Giuseppe Carleo and Matthias Troyer.
\newblock Solving the quantum many-body problem with artificial neural
  networks.
\newblock {\em Science}, 355(6325):602--606, 2017.

\bibitem{torlai2018neural}
Giacomo Torlai, Guglielmo Mazzola, Juan Carrasquilla, Matthias Troyer, Roger
  Melko, and Giuseppe Carleo.
\newblock Neural-network quantum state tomography.
\newblock {\em Nature Physics}, 14(5):447--450, 2018.

\bibitem{rupp2012molecular}
Matthias Rupp, Alexandre Tkatchenko, Klaus-Robert M{\"u}ller, and O~Anatole
  Von~Lilienfeld.
\newblock Fast and accurate modeling of molecular atomization energies with
  machine learning.
\newblock {\em Physical Review Letters}, 108(5):058301, 2012.

\bibitem{ETH1}
Raban Iten, Tony Metger, Henrik Wilming, L\'idia del Rio, and Renato Renner.
\newblock Discovering physical concepts with neural networks.
\newblock {\em Physical Review Letters}, 124:010508, 2020.

\bibitem{ETH2}
Hendrik~Poulsen Nautrup, Tony Metger, Raban Iten, Sofiene Jerbi, Lea~M.
  Trenkwalder, Henrik Wilming, Hans~J. Briegel, and Renato Renner.
\newblock Operationally meaningful representations of physical systems in
  neural networks.
\newblock {\em arXiv preprint arXiv:2001.00593}, 2020.

\bibitem{gurvits2003classical}
Leonid Gurvits.
\newblock Classical deterministic complexity of {Edmonds'} problem and quantum
  entanglement.
\newblock In {\em Proceedings of the thirty-fifth annual ACM symposium on
  Theory of computing}, pages 10--19, 2003.

\bibitem{higgins2017BVAE}
Irina Higgins, Loic Matthey, Arka Pal, Christopher Burgess, Xavier Glorot,
  Matthew Botvinick, Shakir Mohamed, and Alexander Lerchner.
\newblock beta-vae: Learning basic visual concepts with a constrained
  variational framework.
\newblock {\em Iclr}, 2(5):6, 2017.

\bibitem{Perez_PPT}
Asher Peres.
\newblock Separability criterion for density matrices.
\newblock {\em Phys. Rev. Lett.}, 77:1413--1415, Aug 1996.

\bibitem{HORODECKI_PPT}
Michał Horodecki, Paweł Horodecki, and Ryszard Horodecki.
\newblock Separability of mixed states: necessary and sufficient conditions.
\newblock {\em Physics Letters A}, 223(1):1 -- 8, 1996.

\bibitem{qutip2012}
J.R. Johansson, P.D. Nation, and Franco Nori.
\newblock Qutip: An open-source python framework for the dynamics of open
  quantum systems.
\newblock {\em Computer Physics Communications}, 183(8):1760 -- 1772, 2012.

\bibitem{qutip2013}
J.R. Johansson, P.D. Nation, and Franco Nori.
\newblock Qutip 2: A python framework for the dynamics of open quantum systems.
\newblock {\em Computer Physics Communications}, 184(4):1234 -- 1240, 2013.

\bibitem{VAE_Paper}
Diederik~P Kingma and Max Welling.
\newblock Auto-encoding variational bayes.
\newblock {\em arXiv preprint arXiv:1312.6114}, 2013.

\bibitem{kingma_Adam}
Diederik~P Kingma and Jimmy Ba.
\newblock Adam: A method for stochastic optimization.
\newblock {\em arXiv preprint arXiv:1412.6980}, 2014.

\bibitem{chollet2015keras}
Fran\c{c}ois Chollet et~al.
\newblock Keras.
\newblock \url{https://github.com/fchollet/keras}, 2015.

\bibitem{srivastava2014dropout}
Nitish Srivastava, Geoffrey Hinton, Alex Krizhevsky, Ilya Sutskever, and Ruslan
  Salakhutdinov.
\newblock Dropout: a simple way to prevent neural networks from overfitting.
\newblock {\em The journal of machine learning research}, 15(1):1929--1958,
  2014.

\bibitem{hinton2006pretraining}
Geoffrey~E Hinton and Ruslan~R Salakhutdinov.
\newblock Reducing the dimensionality of data with neural networks.
\newblock {\em science}, 313(5786):504--507, 2006.

\end{thebibliography}

\end{document}